\documentclass[12pt,titlepage]{article}
\usepackage{amssymb}
\usepackage{epsfig}
\usepackage{bm}
\font\grande=cmr10 scaled \magstep4
\font\medio=cmr10 scaled \magstep2
\outer\def\beginsection#1\par{\medbreak\bigskip
      \message{#1}\leftline{\bf#1}\nobreak\medskip
\vskip-\parskip
      \noindent}

\newcommand{\eq}{\begin{equation}}
\newcommand{\eqx}{\end{equation}}
\newcommand{\eqn}{\begin{eqnarray}}
\newcommand{\eqnx}{\end{eqnarray}}
\newcommand{\ad}{{a^{\dagger}}}
\newcommand{\fd}{f^{\dagger}}
\newcommand{\Ad}{{A^{\dagger}}}
\newcommand{\Qd}{{Q^{\dagger}}}

\newcommand{\nn}{\nonumber}

\begin{document}
\titlepage

\begin{flushright}
\vspace{15mm}
CERN-PH-TH/2005-269\\
TPJU-9/2005
\end{flushright}
\vspace{15mm}
\begin{center}

\grande{ Planar Quantum Mechanics:  an Intriguing   Supersymmetric Example}

\vspace{15mm}

\large{G. Veneziano}

\vspace{10mm}

 {\sl Theory Division, CERN, CH-1211 Geneva 23, Switzerland }

{\sl and}

{\sl Coll\`ege de France, 11 place M. Berthelot, 75005 Paris, France}
\vspace{10mm}

   \large{J. Wosiek}

   \vspace{10mm}

   {\sl M. Smoluchowski Institute of Physics, Jagellonian University}

{\sl Reymonta 4, 30-059 Krak\'{o}w, Poland}

\end{center}

\centerline{\medio  Abstract}
\vskip 5mm
\noindent
 After setting up a Hamiltonian formulation of  planar  (matrix) quantum mechanics, we illustrate its effectiveness in a non-trivial supersymmetric example. The numerical and analytical study of two sectors of the model,  as a function of 't Hooft's coupling $\lambda$, reveals both a phase transition at $\lambda=1$ (disappearence of the mass gap and discontinuous  jump in  Witten's index) and a new form of strong-weak duality for $\lambda \rightarrow 1/\lambda$.
\vspace{5mm}

\vfill
\begin{flushleft}
CERN-PH-TH/2005-269 \\
TPJU-9/2005\\
December 2005\\
\end{flushleft}

\newpage

\section{Introduction}
More than thirty years after its introduction \cite{'tH} the large-$N$ limit of four-dimensional QCD remains elusive.
It is widely believed that such a limit should capture the most interesting non-perturbative properties of QCD, such as confinement and spontaneous chiral symetry breaking, while neglecting others (e.g. resonance widths and the $U(1)$ anomaly).
It is also believed that, in 't Hooft's limit, QCD should lend itself to an effective description in terms of
a string theory (see e.g. \cite{Polyakov} for  recent developments of the subject) or, perhaps, of a gravitational dual similar to the one enjoyed by ${\cal N} =4$ super-Yang--Mills (SYM) theory through the AdS-CFT correspondence \cite{ADSCFT}.
Indeed, the large-$N$ classification of diagrams according to topology
closely resembles the loop expansion of string theory in terms of surfaces of increasing genus.

One should keep in mind, however, that the connection between large-$N$ and graph-topology
is only proven order-by-order in perturbation theory. Whether the true large-$N$ limit (defined as solving exactly the theory at finite $N$ and {\it then} taking $N$ to infinity) does actually coincide with the non-perturbative solution of a suitably defined ``lowest genus" theory remains to be proven case by case.
Already the Gross-Witten model \cite{GW} and the work of Marinari and Parisi \cite{MP} have taught us that the $N \rightarrow \infty$ limit may not commute with other limits, such as the full resummation of the strong-coupling expansion or  approaching first a phase transition. In QCD itself,  the assumption (now supported by lattice calculations \cite{Giusti}) that the topological susceptibility depends on whether the large-$N$ limit is taken before or after the chiral limit
provides a solution of the $U(1)$ problem \cite{WV}.

In this letter we shall consider a class of planar matrix models in a (possibly new) Hamiltonian formulation. (Large-$N$ literature being very vast, we shall refer to the nice reprint collection \cite{BW}
for the classic papers on the subject).
 We shall then solve a particular supersymmetric case (both numerically and analytically) in a planar (or better lowest-genus) approximation and point out several amusing features of the solution, including a  phase transition  and a non-trivial strong-weak duality in 't Hooft's coupling $\lambda$.

One motivation for this work was to  prepare the ground for checking, in a simpler context,
a recently claimed planar equivalence \cite{ASV} between a supersymmetric ``parent theory" and its
 non-supersymmetric ``daughter" in a particular subsector. Our method (or at least its numerical part)
 should apply without  major modifications
to the latter theory, and therefore such a check should be forthcoming.
Another motivation  came from the recent studies of the supersymmetric Yang-Mills quantum mechanics
in various dimensions \cite{JW}. Although done  at present mainly for SU(2) \cite{CW}
(and partly for SU(3) and SU(4) \cite{JT}) gauge groups,
the goal of these works is to extrapolate eventually towards large $N$
whereby making contact with  M-theory
\cite{BFSS}. Our results here offer the prospect of a substantial
shortcut for the whole program.

The rest of this letter  (see  \cite{Adriano} for more detailed account)  is organized as follows:  we first define a  general class of planar quantum mechanics (PQM) models and specify
some conditions for our method to be applicable. We then focus on
the supersymmetric case and, eventually, on a particular example for which  spectrum  and  main features can be worked out  both numerically and analytically.

\section{Hamiltonian formulation of PQM }
Our formulation of  planar quantum mechanics (PQM) is best done directly in a standard Hamiltonian framework (see also \cite{Raj}).
Let us start by defining a Hilbert space (with states that span it) and operators (acting on it).
The operators will be $N \times N$ destruction and creation matrices:
\eq
^{(k)}M_i^{j}, ~~ ^{(k)}M_i^{j\dagger} \, , \, i, j = 1, \dots N~~ ; ~~  k = 1, \dots N_f \; ,
\eqx
where $i,j$ represent ``colour" indices while $k$ represents a  generalized ``flavour" index. The latter can be used, in a QFT generalization,  as a label
for momenta, polarizations etc. In this paper it will only serve the purpose of   distinguishing  bosonic and fermionic
degrees of freedom. The above operators are assumed to obey  standard (anti) commutation relations.
In familiar notations:

\eq
\left [ ^{(k)}M_i^{j}, ~ ^{(k')}M_l^{m\dagger}\right \} =  \delta_i^m     \delta_l^j   \delta_{k,k'}  \, .
\eqx

The Hilbert space is constructed out of the usual Fock vacuum (annihilated by all $^{(k)}M_i^{j}$)
by acting on it with a {\em single-trace} string of creation operators. This is the first essential difference between general and planar QM. It is of course mimiking the colour structure of the states that
propagate in genus-zero diagrams,  a small subset of all states that are singlets of
a $U(N)$ group acting as:
\eq
M  \rightarrow U  M U^{\dagger} \, .
\eqx
This is of course an enormous simplification: if, for instance,  the index $k$ takes a single (bosonic) value, the states spanning the Hilbert space are just labelled by a single integer, $n$, corresponding to the number of $M^{\dagger}$s in the trace (see below).

The Hamiltonians we shall consider are also single-trace operators and therefore singlets of
$U(N)$, but, of course, they will
include both creation and destruction operators. In order to avoid producing unwanted vacuum diagrams we shall impose  that the Hamiltonian $H$ annihilates the
Fock vacuum. This typically implies (though it is not implied by) normal ordering of the operators appearing in $H$ (not to be confused, of course, with the order in the colour trace).

The Hamiltonian will be a sum of such single-trace operators and will contain a factor $g^{n-2}$
for a term containing $n$ operators. Schematically:
\eq
H = \sum_n c_n  g^{n-2}  Tr (M^n) \, ,
\eqx
where $M^n$ stands for a product of $M$s and $M^{\dagger}$s with a total of $n$ factors.
The ('t Hooft) limit to be consider is, as usual, $N \rightarrow \infty$
with the 't Hooft coupling $\lambda \equiv g^2 N$ kept fixed.

When such a Hamiltonian acts on a generic single-trace state it will not  give, generically,
another single-trace state. However,  whenever it does not, one gets subleading terms in the large-$N$ limit. If we discard such terms we have a closed system and the matrix elements of the Hamiltonian turn out to be functions of $\lambda$ alone:  we simply have a well-defined Hamiltonian to diagonalize in the single-trace Hilbert space.

The final ingredient of our approach is to introduce a cut-off $B \equiv n_{max} $ in the occupation number
thus reducing the problem to one that can be managed numerically \cite{JW} . Eventually, by increasing $B$, we can check whether the lowest eigenstates
and eigenvalues converge to some finite limit. As we shall see, this will be the case in a simple toy model where the method gives very interesting indications
of the dependence of the spectrum from $\lambda$. In turn, the numerical results will suggest properties that we shall be able to derive analytically.

\section{A class of supersymmetric matrix models}

We will now specialize to the case in which there is just one  bosonic and one fermionic
matrix, denoted, respectively, by $a$ and $f$ (plus their Hermitian conjugates).

The class of supersymmetric models that we consider
are a straightforward matrix generalization \cite{MP} of Witten's supersymmetric quantum mechanics
(SQM)\cite{WQM}.
We will assume that  the supersymmetric charges $Q$ and $Q^{\dagger}$ are single-trace
operators that are linear in the fermionic matrices $f$ and $\fd$. Thus:

\eq
Q= Tr (A^{\dagger}f)~,\;\;Q^{\dagger}= Tr(A f^{\dagger})~,\;\;
\eqx
where $A = A(a, \ad)$ represents some function of the bosonic matrices.
We also demand that $Q$ and $Q^{\dagger}$ are nihlpotent. This gives:
\eq
[A^i_j, A^k_l] =0 \, ~, ~~ {\rm for ~ all}~ i, j, k, l \, .
\eqx
In our explicit example we will satisfy this condition trivially by making $A$ ($\Ad$) depend only on
$a$ ($\ad$). We then obtain the supersymmetric Hamiltonian as:
\eq
H=\{Q^{\dagger},Q\} =
(\fd)^i_j~ f^k_l [A^j_i, \Ad^l_k]  + \Ad^i_j~ A^j_i \, .
 \label{Hone}
\eqx
By construction, $H$ commutes with  $Q$ and $Q^{\dagger}$. It also commutes with fermion number,
$F = Tr (\fd~ f)$.
In order to get rid of disconnected diagrams we need the condition:
\eq
A^i_j |0> = 0 \, , ~~ {\rm for~all}~ i, j  \, ,
\eqx
which is again trivially satisfied if  $A$ depends only on
$a$, something that  guarantees that also  $Q$ and $Q^{\dagger}$
annihilate the trivial (empty) Fock state.
Thus, by construction, our model has (at least) one zero-energy state and does not break supersymmetry.

The spectrum of the theory should then consist of a zero-energy sector (providing a certain value of Witten's index) and degenerate massive supermultiplets.
It is easy to show\footnote{One of us (GV) wishes to thank A. Veinshtein for a useful discussion on this issue.}
that, barring unexpected extra symmetries, these supermultiplets should consist of just one boson and one fermion.
Technically, this is a consequence of the fact that the only other operator in the SUSY algebra (besides
$Q$, $Q^{\dagger}$ and $H$), $C \equiv [Q^{\dagger}, Q]$,
satisfies the equation $C^2 = H^2$. Hence, eigenstates can be classified according to whether they carry $C = \pm H = \pm E$ (in amusing analogy with BPS states).  Furthermore, the algebra implies that states with positive (negative) $C$ are annihilated by $Q^{\dagger}$ ($Q$), while they are transformed in a state with opposite $C$ by the other supersymmetric charge. All non-zero-energy levels  must therefore consist of two states  with opposite $C$-parity. For the $F=0$ ($F=1$) sector $C$ is negative (positive) for {\it all} the states but this fails to be the case for higher values of $F$ (see Section 6).

\section{A specific model and its numerical analysis }

We now specialize further our model by taking:
\eq
Q= Tr [f \ad(1+g\ad)] = Tr [f A^{\dagger}],
\;\;\; \Qd=  Tr [\fd (1+g a) a]=  Tr [\fd A],
\eqx
and therefore
\eq
H=H_B+H_F  \, ;
\eqx
\eq
H_B= Tr [\ad a + g(\ad^2 a + \ad a^2) + g^2 \ad^2 a^2] \, ;
\eqx
\eqn
H_F&=& Tr [\fd f + g ( \fd f (\ad+a) + \fd (\ad+a) f) \nonumber \\
& + & g^2 ( \fd a f \ad + \fd a \ad f + \fd f \ad a + \fd \ad f a)] \, .
\eqnx
In most of this paper we shall limit our attention to the $F=0$ and $F=1$ sectors. Some discussion of our
expectations for the $F\ge 2$ will be given at the end of the paper but a detailed analysis is postponed to
further work.

As already anticipated, the planar states in the $F=0$ sector are simply labeled by the integer $n$
corresponding to the number of $\ad$s in the trace. We shall denote the normalized state with $n$ bosonic
quanta by $|0, n \rangle$. Similarly, the generic (single-trace) $F=1$ normalized state will be denoted by
$|1, n-1 \rangle$:  the corresponding creation operator contains one fermionic and $n-1$ bosonic operators. In the free theory ($g=0$) there is a single zero-energy bosonic state, $|0, 0\rangle$, while
$|0, n \rangle$ and $|1, n-1 \rangle$ form a supermultiplet.

Working out the matrix elements of the Hamiltonian is straightforward although tedious
(in particular the normalization factors have to be kept accurately).
Keeping only the leading terms as $N\rightarrow \infty$ we find that the
final result for the matrix elements of $H$ depend only on
 $\lambda \equiv g^2 N $,  and are
given by:

\eq
\langle 0,n|H|0,n \rangle =\left(1+ \lambda (1-\delta_{n1})\right) n \label{h2eq1} \, ,
\eqx
\eq
\langle 0,n+1|H|0,n\rangle=\langle0,n|H|0,n+1\rangle= \sqrt{\lambda}\sqrt{n(n+1)} \, , \label{h2eq2}
\eqx
\eq
\langle 1,n|H|1,n\rangle= (n+1)(1 + \lambda) +  \lambda \, ,
\eqx
\eq
\langle 1,n+1|H|1,n\rangle=\langle1,n|H|1,n+1\rangle=\sqrt{\lambda}(2+n) \, .
\eqx

After introducing a cutoff $B \equiv n_{max}$ we can diagonalize the Hamiltonians in the two sectors and
compute the spectra. Eigenvalues with $E << B$ converge rapidly to some finite values,
except near
$\lambda =1$, where some critical slowdown of the calculation emerges.
Fig. 1 gives the lowest fermionic and bosonic eigenvalues as functions of $\lambda$. Apart from the
trivial bosonic ground state we observe that:
\begin{itemize}
\item There is  excellent boson-fermion degeneracy if we stay away from $\lambda =1$;
\item The cutoff $B$ explicitly breaks supersymmetry, which we expect to recover only at $B= \infty$;
\item The breaking of SUSY allows the supermultiplets to split near $\lambda =1$.
More amusingly, above  $\lambda =1$, the supermultiplets form once more,
but with new partners. The $|0, E_1 \rangle$ state remains unpaired
 (with zero energy),
while $|0, E_{n+2} \rangle$  pairs with $|1, E_n \rangle$ ($n = 0,1, \dots$) rather
than with the small-coupling partner $|1, E_{n+1} \rangle$;
\item Eigenvalues tend to collapse to zero  at $\lambda =1$ as the cutoff is increased.
\item  Some kind of symmetry between strong and weak 't Hooft coupling appears.
\end{itemize}

\begin{figure}[tbp]
\begin{center}
\epsfig{width=8cm,file=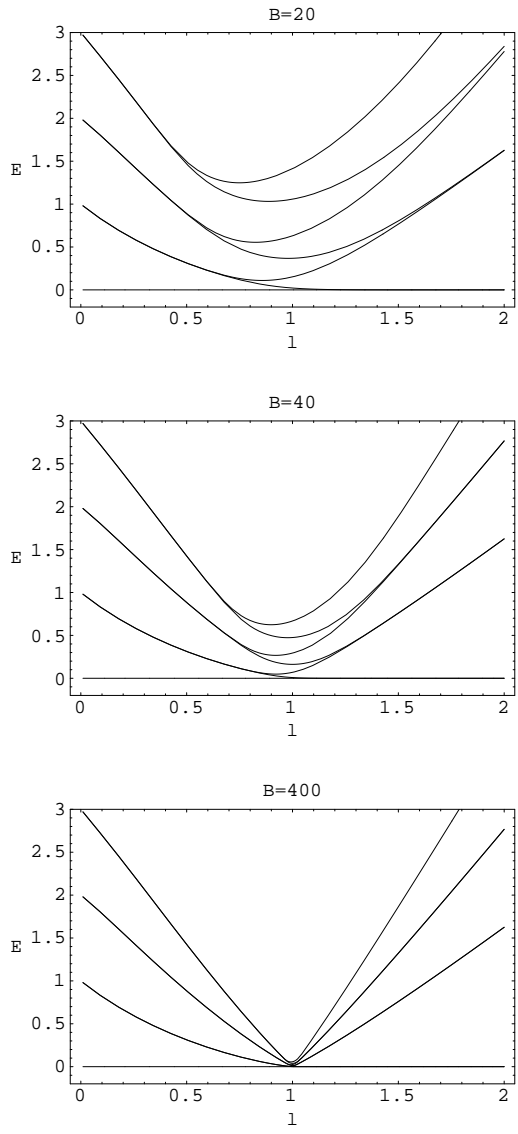}
\end{center}
\vskip-4mm
\caption{Lowest bosonic and fermionic levels as functions of $\lambda$ for different cutoffs}
\label{fig:susyf01}
\end{figure}

Obviously, the behaviour near $\lambda=1$ is strongly suggestive of a phase transition (if the cutoff is removed). A rather shocking way of showing this is to plot the Witten index and partition function (restricted to the two sectors we have considered):
\eq
W(\beta,\lambda) \equiv Tr  \left( (-1)^F e^{-\beta H} \right) ~~~, ~~~ Z(\beta,\lambda) \equiv Tr  \left( e^{-\beta H} \right)
\eqx
The results are shown in Fig. 2. The sudden jump  by one unit  in $W(\beta,\lambda) $ around $\lambda =1$ is
quite spectacular.
The standard, thermal partition function shows even more dramatic singularity at $\lambda=1$. As expected, the large
cutoff and large $\beta$ limits do not commute. Our numerical results suggest that the large cutoff limit at fixed
$\beta$ reveals a singularity of $Z$ at $\lambda=1$ which tends to a step-function  (i.e. as for $W$) plus
a kind of ``delta function" as $\beta\rightarrow\infty$.

In order to understand better these numerical results and what they mean at infinite cutoff, we now resort to some analytic methods.

\begin{figure}[tbp]
\begin{center}
\epsfig{width=14cm,height=6cm,file=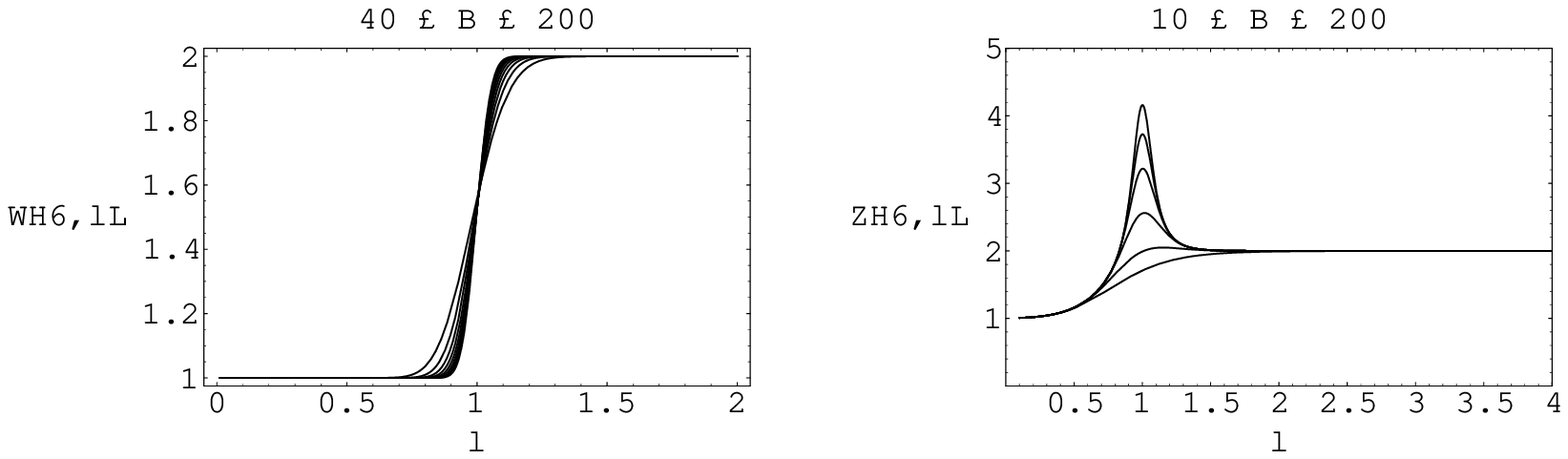}
\end{center}
\vskip-4mm
\caption{$\lambda$ dependence of the Witten index and the partition functiom, at $\beta = 6$
for different cutoffs.}
\label{fig:susyf01}
\end{figure}

\section{Analytic solution}

Let us introduce
 new ``composite" creation and annihilation operators for single trace states:
\eq
\ad_n \,  (a_n) \,  {\rm creates} \,   {\rm (annihilates)} \, |0,n\rangle ~~ ; ~~ \fd_n \, , ( f_n)  \,  {\rm creates} \,   {\rm (annihilates)} \, |1,n-1\rangle \, ,
\eqx
that (anti)commute as usual. Introducing for convenience $b \equiv \sqrt{\lambda}$,
it is  easy to show that:
\eq
 H^{(F=0)} = \ad_1 a_1 + \sum_{n=2}^{\infty} n(1+ b^2) \ad_n a_n +
(\sum_{n=1}^{\infty} b \sqrt{n(n+1)} \ad_n a_{n+1} + h.c.)
\eqx
and
\eq
H^{(F=1)} =
 \sum_{n=1}^{\infty}[n+( n+1) b^2] \fd_n f_n +
(\sum_{n=1}^{\infty} b (n+1) \fd_n f_{n+1} + h.c.)
\label{HF}
\eqx
We can also construct the SUSY charges as:
\eq
 Q =  \ad_1 f_1 + \sum_{n=1}^{\infty} \sqrt{ n+1}\ad_{n+1} (f_{n+1} + b  f_n) \, ,
 \eqx
(and similarly for $\Qd$)
and check  that $\{Q,\Qd\} = H$.

In the $F=0$ sector, besides the trivial vacuum, $|0\rangle_1$, one can formally construct a second state
annihilated by  $H$:
\eq
|0\rangle_2 = \sum_{n=1}^{\infty} \left(\frac{-1}{b}\right)^{n} \frac{\ad_n}{\sqrt{n} }|0\rangle_1 \, .
\label{vac2}
\eqx Clearly  its norm is only finite at $b >1$ explaining
why there is no such a zero-energy state below $b =1$. Using
the formula given above for $\Qd$ one can also check that $\Qd
|0\rangle_2 = 0$.

We now come to the derivation of a new sort of strong-weak-$\lambda$ duality which surprisingly exists in this model.
Using (\ref{HF}) for both $b$ and $1/b$  we find immediately:
\eq
 b H^{(F=1)}(1/b) = \frac{1}{b}  H^{(F=1)}(b)  + (1/b -b) \, ,
\eqx
since $\sum_n \fd_n f_n =1$ in this sector.
 Because of SUSY
it must also work in the $F=0$ sector. In terms of eigenenergies, the duality relations read:
\eq
 b \left(E^{(F=1)}_n(1/b) +1\right)   = \frac{1}{b} \left( E^{(F=1)}_n(b)   +1\right)  \, ; \,
 b \left(E^{(F=0)}_n(1/b) +1\right)  = \frac{1}{b} \left( E^{(F=0)}_{n+1}(b)  +1\right)  \, \, .
 \label{DR}
\eqx
Notice that, due to the existence of the second vacuum for $b>1$,
states, in the $F=0$ sector,  whose
energies are related by duality, {\it do not}  have the same $n$.
These duality relations are nicely satisfied by our numerical eigenvalues computed at large cutoff (see Table 1).

\begin{table}[tbp]
\begin{center}
\begin{tabular}{cccccc} \hline\hline
  \multicolumn{2}{c}{b}       &     $ 1/2 $  & $2$           &  $3/4$  & $4/3$   \\  \hline
   $n$         &    $B$       &              &               &         &         \\  \hline
               &    $10$      &   $3.18818$  &  $3.19736$    &  $1.68608$       &  $1.91640$       \\
        $1$    &    $20$      &   $3.18808$  &  $3.18808$    &  $1.67497$       &  $1.68238$       \\
               &    $30$      &   $3.18808$  &  $3.18808$    &  $1.67488$       &  $1.67499$       \\
  \multicolumn{2}{c}{exact}   &\multicolumn{2}{c}{3.18807663} &  \multicolumn{2}{c}{1.67488116}    \\
               &    $20$      &   $6.45527$  &  $6.45637$    &  $3.09853$       &  $3.45017$       \\
        $3$    &    $40$      &   $6.45524$  &  $6.45524$    &  $2.97193$       &  $2.97543$       \\
               &    $60$      &   $6.45524$  &  $6.45524$    &  $2.97177$       &  $2.97177$       \\
  \multicolumn{2}{c}{exact}   &\multicolumn{2}{c}{6.45523985} &  \multicolumn{2}{c}{2.97177086}    \\
               &    $30$      &   $9.49504$  &  $9.49513$    &  $4.46674$       &  $4.84625$       \\
        $5$    &    $50$      &   $9.49503$  &  $9.49503$    &  $4.19908$       &  $4.21714$       \\
               &    $70$      &   $9.49503$  &  $9.49503$    &  $4.19672$       &  $4.19675$       \\
  \multicolumn{2}{c}{exact}   &\multicolumn{2}{c}{9.49503451} &  \multicolumn{2}{c}{4.19671404}    \\
  \hline\hline
\end{tabular}
\end{center}
\caption{The cutoff dependence of three eigenenergies at two pairs of $b$ values related by
the duality described in the text. Exact results (c.f. below) are identical for dual
partners.}
\label{allFs}
\end{table}

A consequence of duality is that, for all levels, the left and right derivatives of $E$ at $\lambda=1$ ($E'_<, E'_>$)
should satisfy :
\eq
E'_> + E'_<  =1/2 \,
\eqx
a relation that has also been checked numerically.

We finally turn to an analytic  determination of the massive spectrum. To this purpose
it is convenient to rewrite  $H$ in the $F=0$  subspace as:
\eqn
H = \sum_{n=1}^{\infty} B^{\dagger}_n B_n\; , \;\;\;
B_n&=&\sqrt{n} a_n + b \sqrt{n+1} a_{n+1} \, , \nonumber \label{htwo}
\eqnx
and to introduce new states:
\eqn
  |B_n\rangle \equiv B_n^{\dagger}|0\rangle&=&\sqrt{n}|n\rangle+b\sqrt{n+1}|n+1\rangle \, .
\eqnx
These states are not orthonormal, nevertheless they form a complete set and this suffices
 for our construction.
The action of  a ``reduced"  Hamiltonian,  $\bar{H} \equiv
( H - b^2 ) / b $, on the $|B_n>$ states is very simple:
\eqn
\bar{H} |B_n\rangle=  n |B_{n-1}\rangle +
 n \left(b+\frac{1}{b}\right)|B_n\rangle +
(n+1)|B_{n+1}\rangle,\;\;n=2,\dots. \label{htwonb}
\eqnx
with two exceptions at $n=0,1$ for which:
\eqn
\bar{H} |B_0\rangle = - b |B_0\rangle + |B_1\rangle, \;\;\;\;\;\;\;
\bar{H} |B_1\rangle = \left(b+\frac{1}{b}\right)|B_1\rangle +
2|B_{2}\rangle. \nn
\eqnx
The simplicity of (\ref{htwonb}) allows us to map the eigenproblem of $\bar{H}$ into a simple differential
equation.
Let us expand any generic eigenstate of $\bar{H}$ into the $|B_n\rangle$ basis
and associate with it a function of one variable $x$:
\eqn
|\psi\rangle = \sum_{n=0}^{\infty} c_n |B_n\rangle \;,\;\;\;\;\;\;\leftrightarrow\;\;\;\;\;\;
f(x)=\sum_{n=0}^{\infty} c_n x^n \, , \label{spsi}
\eqnx
which is in fact a generating function for the $\{ c_n \}$ coefficients.

It is then easy to see that the eigenequation for $|\psi\rangle$ maps into the following first order
differential equation for $f(x)$:
\eq
 w(x)f'(x) + x f(x) - b f(0) - f'(0) = \epsilon f(x),  \label{hx}
\eqx
where  $w(x)=(x+b)(x+1/b)$, and $\epsilon$ is the eigenvalue of $\bar{H}$ ($E = b(\epsilon+b)$).
The solution of (\ref{hx}) is straightforward:
\eq
f(x)=g(x) \int_{x_0}^x \frac{b f(0)+f'(0) }{w(x')g(x')}dx', \label{fsol} \;
\eqx
where:
\eqn
g(x)=(x+b)^{-\alpha} (x+1/b)^{\alpha-1},\;\;\;\alpha=\frac{\epsilon+b}{b-1/b}~, ~ E = \alpha(b^2-1) \, ,
\eqnx
 is a solution of the homogenous equation and
 $x_0$ is to be determined by some boundary condition.

However, since the inhomogeneous term is given by $b f(0)+f'(0)$, there is an additional consistency
condition, namely the solution and its derivative,
when taken at $x=0$, must reproduce $f(0)$ and $f'(0)$ again. This leads to the relations
\eq
{\rm either}\;\;\;(b g(0)+g'(0) )=0,\;\;\;\;{\rm or}\;\;\;\;
\int_{x_0}^0 dx (x+b)^{\alpha-1}(x+1/b)^{-\alpha}=0.
\label{eigeq}
\eqx
The first condition gives $\alpha=0$, hence it can only lead to zero-energy states. Thus the massive
spectrum follows from
the second condition (\ref{eigeq}).
 Consistency with (\ref{vac2}) fixes $x_0=-1/b$ for $b >1$ and $x_0=- b$ for $b < 1$. In either case
one should use the analytic continuations of Eq.(\ref{fsol}) in order to solve (\ref{eigeq}).

Once this is done,  our solution can be written in terms of the standard hypergeometric function $F \equiv  _2F_1$ as
\eqn
f(x)&=& \frac{1}{\alpha}\;\frac{1}{x+1/b}\;  F(1,\alpha;1+\alpha;\frac{x+b}{x+1/b}),\;\;\; b < 1, \\
f(x)&=& \frac{1}{1-\alpha}\;\frac{1}{x+b}\; F(1,1-\alpha;2-\alpha;\frac{x+1/b}{x+b}),\;\;\; b > 1,
\eqnx
and provides the generating functions for the expansion coefficients $\{c_n\}$ of the arbitrary eigenstate
into the $|B_n\rangle$ basis. As one crosscheck examine the $b>1$ solution at $\alpha=0$ to find that it
indeed generates the second massless state, Eq.(\ref{vac2}). On the other hand, similar attempt for the $b<1$
solution fails -- there is no such state in the weak coupling regime.

To summarize, after some trivial change of integration variable, the non-zero-energy levels of the $F=0$
(and thus by SUSY also of the $F=1$) sector are given by the roots in $\alpha$ of the following equations:
\eq
\int_0^{1/\lambda} dx (1-x)^{-1} x^{-\alpha} = 0  \, ,  \; \; (\lambda >1)  \;\; ; \;\;
\int_0^{\lambda} dx (1-x)^{-1} x^{\alpha -1} = 0  \, ,  \; \; (\lambda < 1) \, .
\label{incB}
\eqx

Solving these equations indeed reveals a series of discrete zeros, $\alpha_n > (<) 0$ for $b > (<) 1$ which nicely confirm
the eigenvalues $E_n = \alpha_n (\lambda -1)$  computed numerically in the previous section.
One immediately  checks that, for two values of $\lambda$ related by $\lambda \rightarrow 1/\lambda$,
the solutions for $\alpha$ are
connected by $\alpha \rightarrow 1- \alpha$, insuring the duality relations (\ref{DR}) among
the corresponding eigenvalues.

At this point one can study the flow of the eigenvalues in various situations, e.g. at very weak
(and thus by duality also at very strong) coupling.
More interesting is the behaviour of the eigenvalues near the critical point at $\lambda =1$.
The Beta-functions in (\ref{incB}) can be related to
 $F (\alpha, 1, \alpha +1 ; \lambda)$ and $F(1- \alpha, 1, 2- \alpha ; 1/\lambda)$ for $\lambda <1$ and $\lambda >1$, respectively.
>From the known asymptotic behaviour of $F$ (as its last argument approaches 1), we easily get
the  approximate eigenvalues around  $\lambda =1$ in the form:
\eqn
 \lambda \rightarrow 1^- &:&  ~~ E_n  = (-\alpha_n) (1 - \lambda) \;  ,
\nn \\
\lambda \rightarrow 1^+ &:&~~E_{n+1}  = (1- \alpha_n) (\lambda -1) \;, \;
n = 0, 1, 2, \dots \, ,
\nn \\
\psi(\alpha_n) &+& \gamma + log(|1-\lambda|) + O\left( |1-\lambda| log(|1-\lambda|)\right) = 0 \, ,
\eqnx
where $\psi$ is the logarithmic derivative of the $\Gamma$-function and $\gamma = 0.5772..$
is the Euler-Mascheroni  constant.
These formulas obey the duality relations (\ref{DR}). They also show  the non-analytic way  the various
levels collapse to zero energy at the critical point. In particular, as $\lambda \rightarrow 1^-$, the first eigenvalue
 approaches zero as $- (1-\lambda)~ log^{-1}(1-\lambda)$,  i.e. with vanishing first  --and  infinite second-- derivative.

The above formulae also allow a quantitative study of the free energy of the model near the phase transition,
which appears to be
stronger than in the Gross-Witten model.
We have also determined numerically the first few
zeroes at generic values of $\lambda$,  and found perfect agreement with the large cutoff limits of the
numerical eigenvalues. The only slightly difficult comparison occurs just around the phase transition
where convergence (as one increases the cutoff) undergoes a critical slowdown.

\section{Remarks about the $F \ge 2$ sectors}
In principle our analysis can be extended in a straightforward way to higher fermion-number sectors.
In practice,  calculation of the Hamiltonian in those sectors becomes quickly cumbersome.
There are at least two reasons why it would be  worthwhile making such an extension.

Firstly, one would like to check whether the $F=0,1$ sectors completely determine the structure of the phase transition at $\lambda=1$.  This would depend on how eigenvalues in
the higher-$F$ sectors behave near $\lambda=1$ and in particular on whether
there are discontinuous jumps of Witten's index also in those sectors.

The second reason is that we may expect qualitatively new phenomena to occur when we consider
$F \ge 2$ sectors.
The fact that eigenstates in the $F=0$ and $F=1$ sectors pair nicely without involving, say, $F=2$ states
can be argued on the basis of simply counting the former states at weak coupling.
However, when we go to higher $F$, states are  typically highly degenerate at zero coupling.
The counting is relatively easy and is summarized in a kind of ``Chew-Frautschi"
plot in Fig. 3, where non-degenerate states are marked with a full circle while degenerate ones
are represented by an open circle showing the degree of degeneracy.

\begin{figure}[tbp]
\begin{center}
\epsfig{width=8cm,file=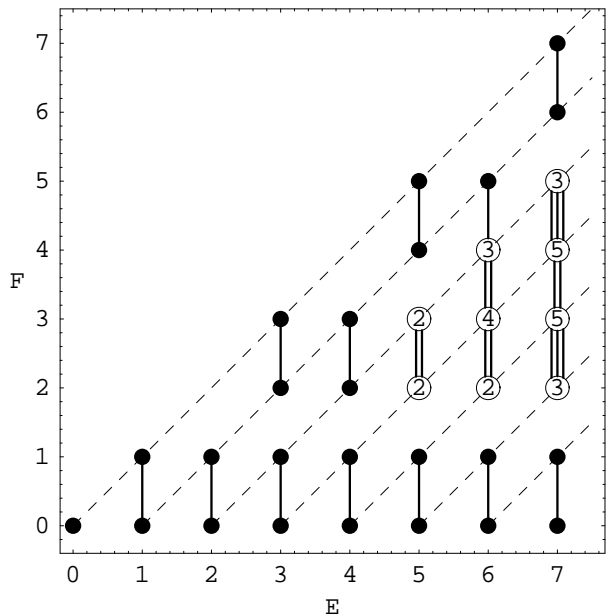}
\end{center}
\vskip-4mm
\caption{Chew-Frautschi-like plot of the  lowest seven levels with their respective degeneracy
and supersymmetry partnership at vanishing $\lambda$.}
\label{fig:cfplot}
\end{figure}
Pairing these states in SUSY doublets (as  $\lambda$ is switched on)  turns out to be  possible, but non trivial
(due to some magic combinatorics! \cite{JW}-\cite{JT}): it  is shown in  Fig. 3 via the vertical segments connecting different circles.  
For instance,  the $E=6$ levels must pair according to the following pattern:
the two $F=2$ states find their SUSY partners in two linear
combinations of the four $F=3$ states. The remaining two $F=3$ states will  match two (linear combinations) of the
three $F=4$ bosons. Finally, the third $F=4$ boson will pair with the single $F=5$ fermion.
One  can also argue that, while most of the above levels have $C/E = (-1)^{F+1}$, two of the
$F=3, E=6$ levels and one of the $F=4, E=6$ levels have  $C/E = (-1)^{F}$.
We are planning to check numerically the low-lying spectra of the $F=2$ and $F=3$ sectors in order to see whether,
 indeed, two of the four $F=3$ states around $E=6$ do split from the two $(F=2,F=3)$
doublets as we turn on $\lambda$. Were this not  the case, would signal some higher symmetry
underlying the model.

\section{Discussion, summary}
In this paper we have presented a new way to tackle, both numerically and analytically, planar
quantum mechanical problems which hopefully represent the large-$N$ limit of matrix models.
Given the ubiquiness of matrix models in theoretical physics (see again \cite{BW}),  it is hard to overestimate the importance of developing powerful techniques for approaching this kind of questions.

Our method is based on a direct Hamiltonian construction of states and operators that are relevant
at lowest genus in a topological expansion of the theory. In principle it should be applicable to any discretization of quantum field theories that allows to compute the planar Hamiltonian in a convenient
basis of vectors.

As an illustration of the method we have considered a supersymmetric quantum mechanics model
and managed to solve for its spectrum, both numerically and analytically, in two fermionic sectors.
Since supersymmetry transformations close within these two sectors, we find, as expected,
boson-fermion degeneracy. To our surprise, however, we also find that, at a critical value of
the 't Hooft coupling, $\lambda=1$, the spectrum looses its mass gap and becomes continuous.
This conclusion is also confirmed by the radically different dependence of the spectrum on the cutoff at $\lambda=1$.
This dependence is indeed characteristic of the scattering  plane-waves \cite{TW}.
On the other side of the critical value the spectrum has once more a mass gap but there is one more
zero-energy bosonic state. In other words the Witten index has jumped by one unit across the phase transition. Furthermore, energy levels on the two sides of the critical point are connected through a non-trivial duality relation. Another amusing property of the model is that, at least within those two sectors,
it can be solved analytically.

Besides generalizing the model to more interesting cases, there are two important directions in which
the model itself deserves further study:
\begin{itemize}
\item Extend calculations to sectors with higher fermion number;
\item Understand the situation at finite (though large) N.
\end{itemize}
We hope to be able to address these issues in a forthcoming paper.

\section*{Acknowledgements}
It is our pleasure to thank P. van Baal, P. Di Vecchia, L. Giusti, V. Kazakov, G. Marchesini,
E. Rabinovici and A. Schwimmer for useful discussions and/or correspondence. This work is partially supported
by the grant of Polish Ministry of Science and Education P03B 024 27 (2004 - 2007).


\begin{thebibliography}{99}
\bibitem{'tH} G. 't Hooft, Nucl. Phys. {\bf B72} (1974) 461; see also G. Veneziano,
Nucl. Phys. {\bf B117} (1976) 519.
\bibitem{Polyakov}
A.M. Polyakov, in  {\em 50 years of Yang-Mills theory}, (G. *'t Hooft, G. editor) World Scientific Publishing Company, Singapore (2004), 311 [ hep-th/0407209].
\bibitem{ADSCFT} J.M. Maldacena, Adv. Theor. Math. Phys. {\bf 2} (1998) 231;  S.S. Gubser, I.R. Klebanov and A.M. Polyakov,   Phys. Lett. {\bf 428} (1998) 105; E. Witten,  Adv. Theor. Math. Phys. {\bf 2} (1998) 253.
\bibitem{GW} D. J. Gross and E. Witten, Phys. Rev. {\bf D21} (1980) 446.
\bibitem{MP} E. Marinari and G. Parisi, Phys. Lett. {\bf B240} (1990) 375.
\bibitem{Giusti}
 L. Del Debbio , L. Giusti and C. Pica, Phys. Rev. Lett. {\bf 94} (2005) 032003
 [hep-th/0407052].
\bibitem{WV} E. Witten, Nucl. Phys. {\bf B156} (1979) 269;
G. Veneziano, Nucl. Phys. {\bf B159} (1979) 213.
\bibitem{BW} {\em The large $N$ expansion in Quantum Field Theory and Statistical Physics} (E. Brezin and S. R. Wadia editors), World Scientific Publishing Company, Singapore (1993).
\bibitem{ASV}
A. Armoni, M. Shifman and G.Veneziano,
Nucl.\ Phys.  {\bf B667}, 170 (2003)
[hep-th/0302163];
  Phys.\ Rev.  {\bf D71}, 045015 (2005)
  [hep-th/0412203].
\bibitem{JW} J. Wosiek, Nucl.\ Phys. {\bf B644 } (2002) 85 \ [hep-th/0203116].
\bibitem{CW} M. Campostrini and J. Wosiek, Nucl.\ Phys. {\bf B703 } (2004) 454 \ [hep-th/0407021].
\bibitem{JT} R. Janik and M. Trzetrzelewski, work in progress.
 \bibitem{BFSS} T. Banks, W. Fischler, S.H. Shenker and  L. Susskind, Phys. Rev. {\bf D55} (1997) 5112
 [hep-th/9610043].
\bibitem{Adriano} G. Veneziano and J. Wosiek, to appear in Adriano Di Giacomo's festschrift (2006).
\bibitem{Raj} C.-W. H. Lee and S. G. Rajeev, Phys.Lett. {\bf B436} (1988) [hep-th/9806019].
\bibitem{WQM} E. Witten, Nucl. Phys. {\bf B185} (1981) 513; {\bf B202} (1983) 253.
\bibitem{TW} M. Trzetrzelewski and J. Wosiek, Acta Phys.\ Polon. {\bf  B35} (2004) 1615,
 [hep-th/0308007].

\end{thebibliography}
\end{document}